# Extending the Mott-Gurney law to one-dimensional nonplanar diodes using point transformations

Allen L. Garner,[1,2,3,a)] N. R. Sree Harsha[1,b)], and Amanda M. Loveless[1]

[1]School of Nuclear Engineering, Purdue University, West Lafayette, Indiana 47907, USA

[2]Department of Agricultural and Biological Engineering, Purdue University, West Lafayette, Indiana 47907, USA

[3]Elmore Family School of Electrical and Computer Engineering, West Lafayette, Indiana 47907, USA

[a)] **Author to whom correspondence should be addressed:** algarner@purdue.edu

[b)] **Present address:** Department of Electrical and Computer Engineering, University of Rochester, Rochester, New York 14627, USA

**Abstract:**

Recent studies have applied variational calculus, conformal mapping, and point transformations to generalize the one-dimensional (1D) space-charge limited current density (SCLCD) and electron emission mechanisms to nonplanar geometries; however, these assessments have focused on extending the Child-Langmuir law (CLL) for SCLCD in vacuum. Since the charge in the diode is independent of coordinate system (i.e., covariant), we apply bijective point transformations to extend the Mott-Gurney law (MGL) for the SCLCD in a collisional or semiconductor gap to nonplanar 1D geometries. This yields a modified MGL that replaces the Cartesian gap distance with a canonical gap distance that may be written generally in terms of geometric scale factors that are known for multiple geometries. We tabulate results for common geometries. Such an approach may be applied to any current density, including non-space-charge limited gaps and SCLCD that may fall between the CLL and MGL.

## I. INTRODUCTION

Electron emission plays a pivotal role in devices for numerous applications, including directed energy, high-power vacuum electronics, time-resolved microscopy, and thermionic converters [1-4]. Regardless of the electron emission mechanisms, a limit is ultimately reached above which the current can no longer be increased. This limit is referred to as the space-charge limited current density (SCLCD). Child and Langmuir derived the limit in vacuum for one-dimensional (1D) planar electrodes, as

$$J_{\text{CL}} = \frac{4\sqrt{2}}{9}\varepsilon_0\sqrt{\frac{e}{m}}\frac{V^{3/2}}{D^2}, \tag{1}$$

where $V$ is the electric potential drop across a gap of distance $D$, $e$ is electron charge, $m$ is electron mass, and $\varepsilon_0$ is the permittivity of free space. The classical derivation of the Child-Langmuir law (CLL) $J_{CL}$ assumes the emission of electrons from the cathode with zero initial velocity, so the condition for SCLC coincides to a vanishing electric field over the entire cathode [5]. However, when the initial velocity $v_0$ is nonzero, the electric field at the cathode is nonzero and the virtual cathode, where the electric field vanishes, is located between the cathode and anode [6,7]. The SCLCD for $v \neq 0$, often attributed to Jaffé [8], becomes

$$\frac{J_{\text{Jaffé}}}{J_{\text{CL}}} = \left[\left(\frac{mv_0^2}{2eV}\right)^{1/2} + \left(1 + \frac{mv_0^2}{2eV}\right)^{1/2}\right]^3. \tag{2}$$

While (1) and (2) consider a 1D planar diode, practical devices are more complicated [9], motivating extensions of these vacuum SCLCD to nonplanar [10-28] and multidimensional geometries [29-37]. While computational tools, such as particle-in-cell (PIC) simulations can effectively predict electron emission behavior, it is often valuable to have analytic solutions to use as benchmarks for simple geometries, such as concentric cylinders. This motivated recent work developing theoretical techniques for extending (1) and (2) to nonplanar diodes. One approach used variational calculus (VC) to extremize the current across the gap to derive analytic solutions for concentric cylinders [23], concentric spheres [23], and tip-to-tip and tip-to-plate geometries [26] without a magnetic field. VC was subsequently used to derive the SCLCD for a concentric cylindrical crossed-field diode with orthogonal electric and magnetic fields [27]. A second tehcnique used conformal mapping, which mapped the electric potential for the space-charge limited diode in a planar diode to more complicated goemetries [25,26], to validate the concentric cylinder solution and expand it to more complicated 1D geometries. The third technique used point transformations, which defines a set of canonical coordiantes that reduces an ordinary differential equation to a form that can be solved directly by quadrature [28]. Practically, these canonical coordinates represent local transformations that map the solution from one coordinate system to another coordinate system and can be thought of as bijective point transformations [38]. We can write the SCLCD for an orthogonal geometry in canonical coordiantes as

$$J_{\text{SCLC,vac}} = \frac{4\sqrt{2}}{9}\varepsilon_0\sqrt{\frac{e}{m}}\frac{V^{3/2}}{\mathcal{D}^2}, \tag{3}$$

where $\mathcal{D}$ is the canonical gap distance given by

$$\mathcal{D}_c = R_C(\ln R_C - \ln R_A), \tag{4a}$$

$$\mathcal{D}_s = R_C(R_A - R_C)^2/R_A, \tag{4b}$$

and

$$\mathcal{D}_{t-t} = a\sin^2(\ln[\tan(\eta_A/2)] - \ln[\tan(\eta_C/2)]), \tag{4b}$$

for concentric cylindrical, concentric spherical, and tip-to-tip geometries, respectively, where $R_C$ and $R_A$ are the radii of the cathode and anode radius, respectively, $\eta_C$ and $\eta_A$ are hyperboloids that represent the cathode and anode, respectively, in prolate spheroidal coordinates, and $a$ is the length scaling for tip-to-tip geometry [28]. For a concentric cylindrical geometry, (3) and (4a) are recovered using variational calculus to extremize current [23], conformal mapping to map the electric potential for a space-charge limited gap from a planar geometry to the geometry of interest [26] and Lie point symmetries to define the relevant equations in canonical coordinates [28]. We can also leverage the forms of the equations for SCLCD for geometries that exhibit curvilinear flow from conformal mapping by applying the relevant canonical gap distance to (3) [39]. Moreover, Harsha *et al.* used Lie point symmetries to extend (3) to write the SCLCD with nonzero initial velocity $v_0$ for any 1D geometry as

$$J_{\text{SCLC,vac},v_0 \neq 0} = \frac{4\sqrt{2}}{9}\varepsilon_0\sqrt{\frac{e}{m}}\frac{V^{3/2}}{\mathcal{D}^2}\left[\left(\frac{mv_0^2}{2eV}\right)^{1/2} + \left(1 + \frac{mv_0^2}{2eV}\right)^{1/2}\right]^3, \tag{5}$$

where we obtain $\mathcal{D}$ from (3) or derive it for the appropriate geometry (e.g., curvilinear coordinates as described immediately above) and the term in the square brackets is the Jaffé correction to (3) for nonzero initial velocity [28].

While these approaches elucidate the meaning of 1D and provide closed-form solutions to benchmark PIC for simple geometries, they are limited to vacuum. Practical devices may operate at imperfect vacuum due to impurities or leakage, which would invalidate the assumptions of (1)-(5). For non-vacuum devices, one may operate at sufficiently high pressures that electron emission transitions from vacuum to a collisional state. The Mott-Gurney law (MGL), given by

$$J_{\text{MG}} = \frac{9}{8}\mu\varepsilon\frac{V^2}{D^3}, \tag{6}$$

where $\mu$ is the electron mobility and $\varepsilon$ is the permittivity of the material (which may be $\varepsilon_0$ in a gas, but will not be, in general, for a semiconductor material or a liquid), defines the SCLCD in a semiconductor gap [40]. This has grown in importance with increased studies in semiconductor development, including flexible sensors [41], memory devices [42], solar cells [43], and capacitors [44]. Understanding the behavior of (6) is also important for estimating charge-mobility in these devices [45,46]. The MGL may also define the SCLCD in a liquid or a combined liquid/gas substance [47]. For gases, the condition can become more nuanced for non-vacuum pressures. Theories demonstrated that multiple electron emission mechanisms and space-charge limits may contribute to the observed current under different operating conditions [48,49]. For instance, under practical conditions at non-vacuum pressures, one may operate near a condition where neither (1) nor (6) exactly defines the space-charge limit. Wang *et al.* showed experimentally that a submicron device at atmospheric pressure could undergo breakdown for a condition near a third-order nexus between the Fowler-Nordheim equation for field emission, the CLL, and the MGL based on matching these equations [50]. Breen and Garner derived an exact solution for the SCLCD for a collisional gap that recovered the vacuum solutions (i.e., CLL and Jaffé solution) in vacuum, the MGL for a fully collisional gap with $v_0 = 0$, and a MGL-equivalent to the Jaffé solution for $v_0 \neq 0$ as [51]

$$J_{\text{BG}} = J_{\text{MG}}\left(\frac{1}{2} - \frac{2}{3}\zeta^2 + \frac{1}{2}\sqrt{1 - \frac{8}{3}\zeta^2 + \frac{64}{27}\zeta^3 - \frac{16}{27}\zeta^4}\right), \tag{7}$$

where

$$\zeta = \frac{mDv\nu}{eV} = \frac{Dv}{V\mu} \qquad (8)$$

and $\nu = e/(m\mu)$ is the collision frequency [52,53]. Some studies have also extended theoretical studies of the MG law to include traps for the Mark-Helfrich (MH) equation [54], nonplanar (e.g., cylinder and spheres) geometries [55], multiple dimensions [56], quantum effects [57], and linkages between the MG and MH equations with field emission [58].

This paper takes the next step in this assessment by extending the MG law to nonplanar diodes by applying bijective point transformations to the covariance of charge in a gap. This differs from our recent studies [39,59], which applied the canonical gap distance directly to the emission and SCLCD equations (but only specifically applied the correction to the vacuum condition) [39] and derived an effective canonical gap distance for a multidimensional collisional diode [59]. Instead, we focus here on applying bijective point transformations directly to the MG law.

Section II reformulates the traditional MG law for a planar diode to generalize the cathode position to facilitate application of bijective transformations [25, 28, 39]. Section III summarizes the mathematics of bijective (one-to-one and onto) point transformations and shows how applying the coordinate system invariance of charge yields a general technique to solve for the SCLC in complicated, nonplanar geometries. We then apply point transformations to multiple geometries previously assessed for vacuum using conformal mapping. We make concluding remarks in Sec. IV.

## II. DERIVATION OF THE MGL IN A PLANAR DIODE

The first step to applying bijective point transformations of a 1D planar diode is to summarize the derivation of the MGL in a planar diode. We write Poisson's equation in 1D Cartesian coordinates as

$$\frac{dF}{dx} = \frac{\rho(x)}{\varepsilon}, \qquad (9)$$

where $F$ is the electric field, $x$ is the spatial coordinate in the direction of the electric field, and $\rho(x)$ is the charge density. From electron continuity, we obtain

$$J = env = \rho(x)F(x)\mu, \tag{10}$$

where $J$ is the current density and $n$ is the electron number density. Note that (10) assumes $v = F(x)\mu$, which is the standard assumption in the classical derivation of the MGL [40]. Combining (9) and (10) yields

$$\frac{dF}{dx} = \frac{j}{\mu\varepsilon F(x)}. \tag{11}$$

The standard definition of the MGL in a planar geometry assumes the cathode at $x = 0$ and the anode at $x = D$; however, following the derivation of Harsha and Garner in vacuum [25], we define the cathode and anode at $x = x_C$ and at $x = x_A$, respectively, to facilitate direct translations to other coordinate systems where defining a location at zero would create a singularity (e.g., concentric cylinders). Applying separation of variables to (10) and integrating from the cathode at $x = x_C$ to a general position $x$ gives

$$F^2 - F_C^2 = \frac{2j}{\mu\varepsilon}(x - x_c), \tag{12}$$

where $F$ is the electric field at $x$ and $F_C$ is the electric field at the cathode. Defining $F_C = 0$ for the space-charge limited condition with zero initial velocity [40,55,59] and $F = d\phi/dx$ gives

$$\frac{d\phi}{dx} = \left[\frac{2J_{MG}}{\mu\varepsilon}(x - x_c)\right]^{1/2}, \tag{13}$$

where we have substituted $J = J_{MG}$ since we are specifically considering the space-charge limited condition. Integrating (13) from $x = x_c$ to $x = x_A$ and rearranging to solve for $J_{MG}$ yields

$$J_{MG} = \frac{9}{8}\frac{V^2\mu\varepsilon}{(x_A - x_C)^3}, \tag{14}$$

which may be written in the more familiar form of (5) for a planar diode by setting $D = x_A - x_c$. Note that we must define nonzero locations for the cathode and anode to avoid singularities in position when transferring to other coordinate systems [25].

### III. BIJECTIVE POINT TRANSFORMATIONS

We next consider a 1D diode with the cathode located at $q_c$ and the anode located at $q_A$ in an orthogonal coordinate system. We map the space between an orthogonal diode described by a Riemannian metric $ds^2 = h_1^2 dq_1^2 + h_2^2 dq_2^2 + h_3^2 dq_3^2$ bijectively onto a canonical metric given by $ds^2 = d\zeta_1^2 + d\zeta_2^2 + d\zeta_3^3$. Covariance of the total charge in the gap requires the coordinates be related by [39]

$$\frac{d\zeta_1}{dq_1} = \frac{h_1}{h_2 h_3}, \quad (15a)$$

$$\frac{d\zeta_2}{dq_2} = \frac{h_2}{h_3 h_1}, \quad (15b)$$

and

$$\frac{d\zeta_3}{dq_3} = \frac{h_3}{h_2 h_1}. \quad (15c)$$

The canonical gap distance is given by

$$\mathcal{D} = h_2 h_3 \int_{q_C}^{q_A} \frac{h_1}{h_2 h_3} dq_1. \quad (16)$$

The metric in canonical coordinates given by $ds^2 = d\zeta_1^2 + d\zeta_2^2 + d\zeta_3^3$ resembles the metric tensor in Cartesian coordinates given by $ds^2 = dx^2 + dy^2 + dz^2$. Hence, *any* solution for a current density given in Cartesian coordinates can be modified by replacing the gap distance $D$ with the canonical gap distance $\mathcal{D}$. Thus, we may combine (14) and (16) to write the corresponding SCLCD in a collisional gap in a general orthogonal coordinate system as

$$J_{MG} = \frac{9}{8} \frac{V^2 \mu \varepsilon}{\mathcal{D}^3}, \quad (17)$$

where $\mathcal{D}$ is the canonical gap distance from (16) [28, 38, 39]. Table I gives $\mathcal{D}$ for the geometries in Fig. 1. Details on the transformations to obtain these results are summarized elsewhere [25,26].

While our prior studies extended vacuum SCLCD and the present study addresses the MGL, conditions may also fall in between these two limits under gas and gap conditions that are neither fully vacuum nor fully collisional [51, 59, 60]. Garner and Harsha recently applied a capacitance argument to derive an analytic estimate of the SCLCD under these intermediate conditions [59] that agreed within ~4.5% of the exact solution [51]. We may apply the same argument to replace $D$ with $\mathcal{D}$ for this analytic solution for intermediate collisionality [59].

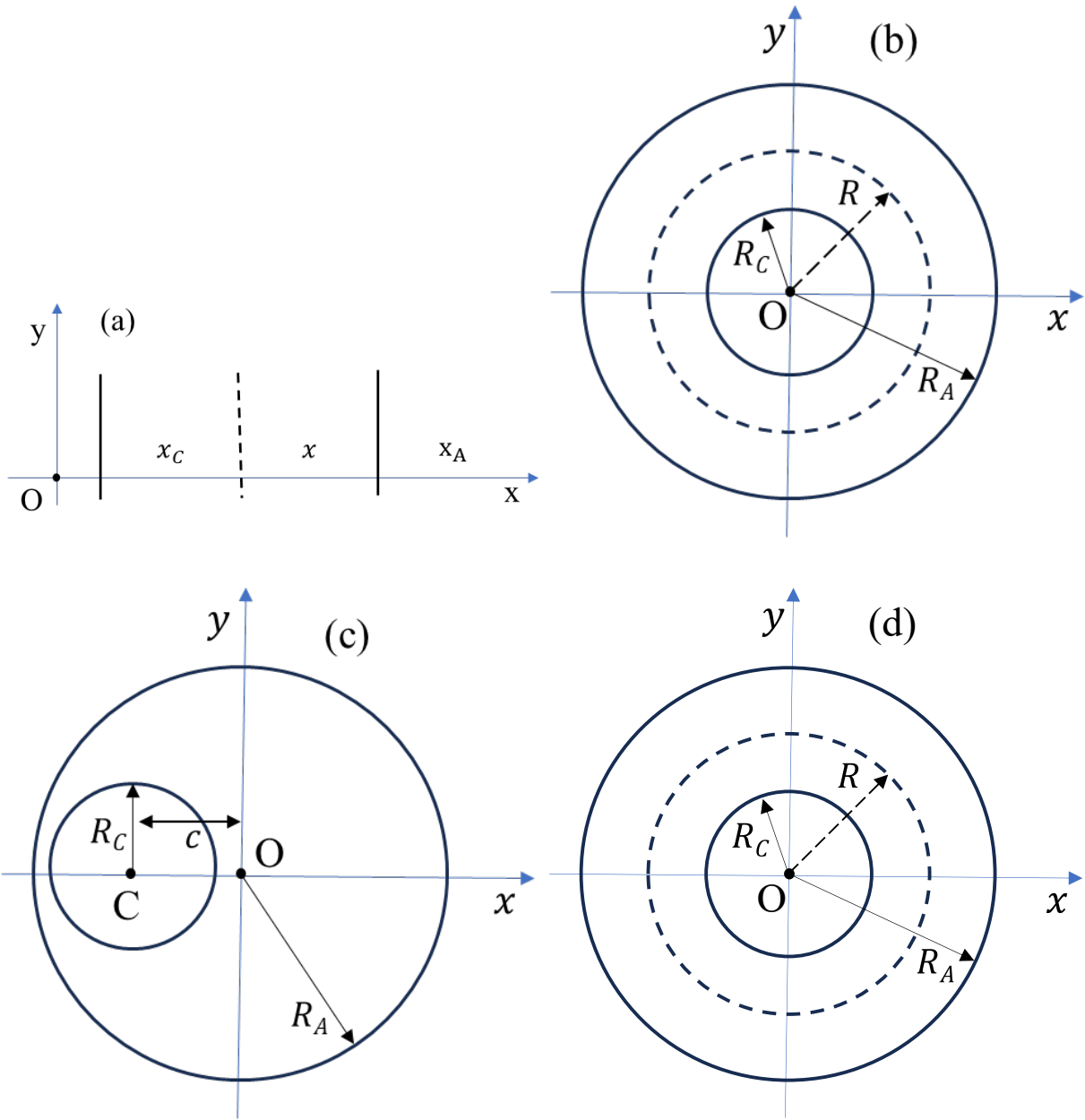

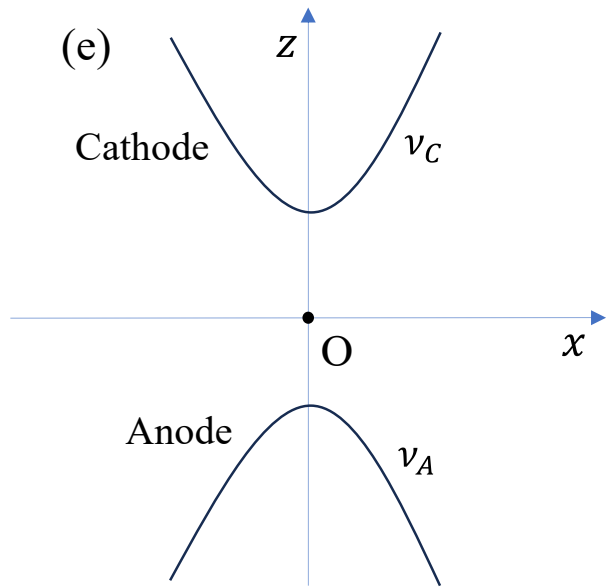
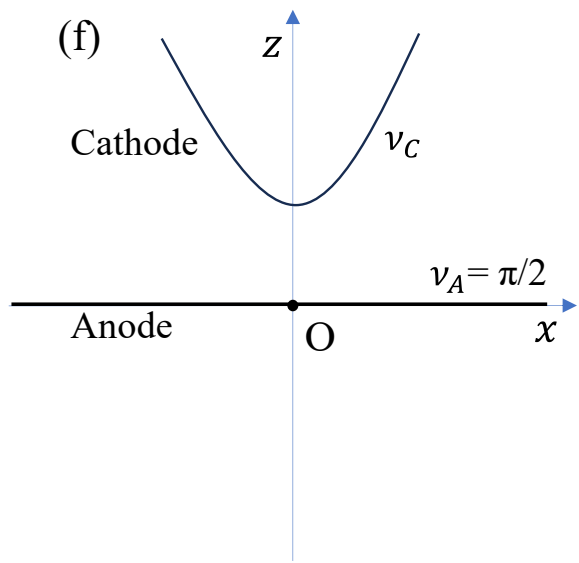
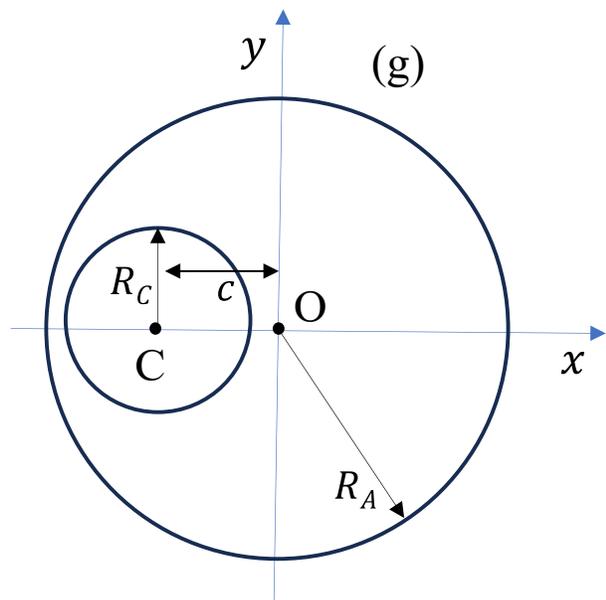
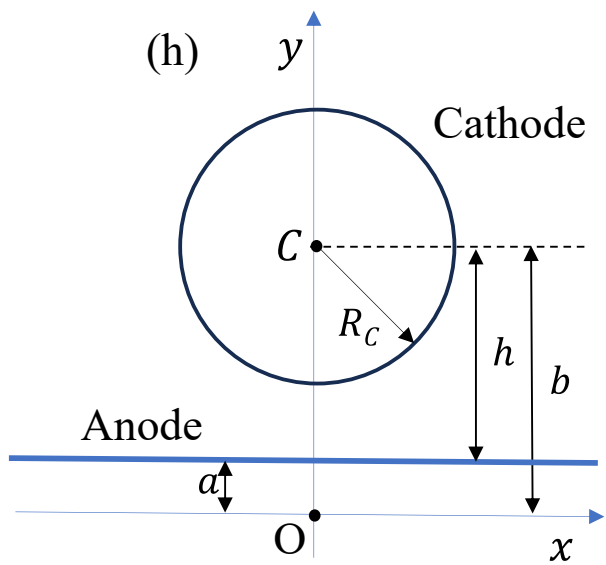

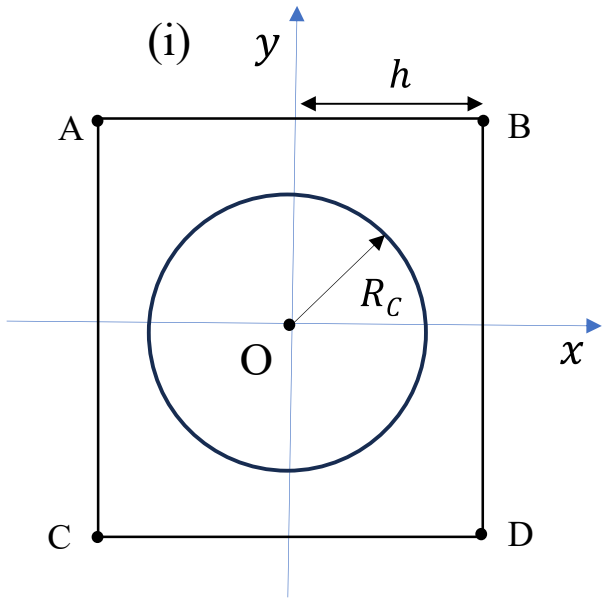
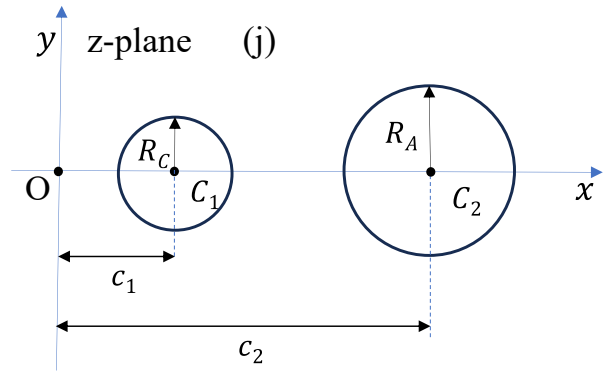
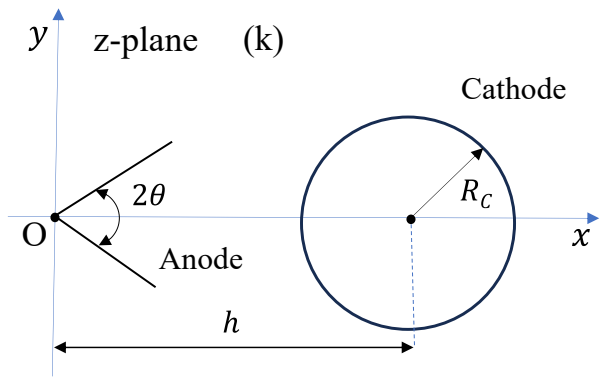
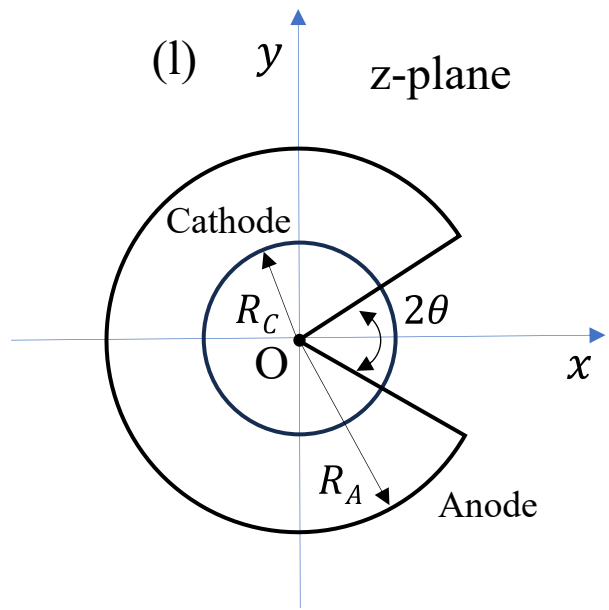

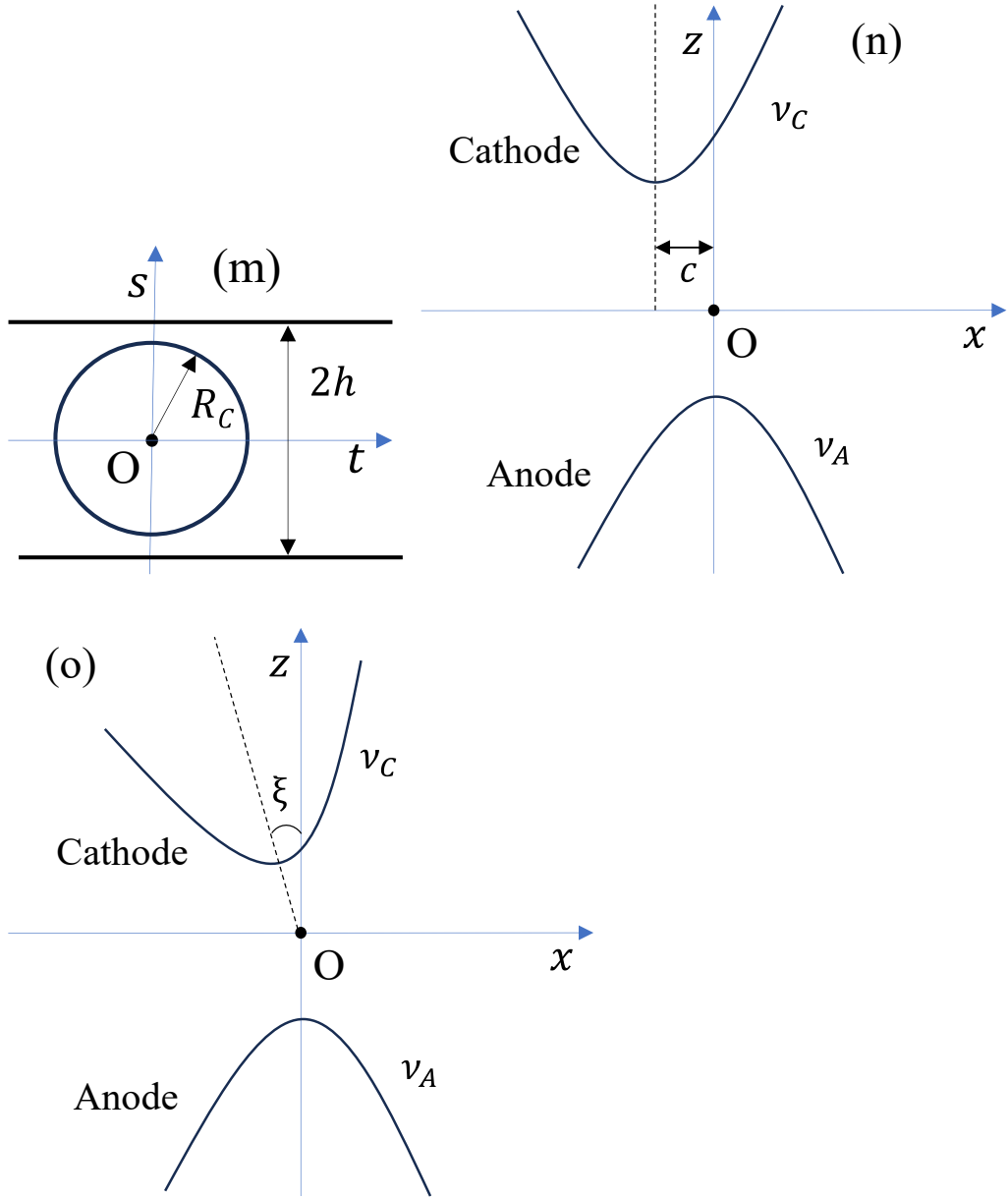

**FIG. 1.** Geometries considered for calculating collisional SCLCD: with canonical gap distances defined in Table I: (a) planar anode and planar cathode; (b) concentric cylinders; (c) eccentric cylinders; (d) concentric spheres; (e) hyperboloid tips; (f) planar anode and hyperboloid tip cathode; (g) eccentric spheres; (h) infinite planar anode and cylindrical cathode; (i) square anode and cylindrical cathode; (j) external cylinders; (k) two intersecting planes anode and cylindrical cathode; (l) slotted cylinders; (m) two plane anode and a cylindrical cathode; (n) tilted hyperboloid tip; (o) misaligned hyperboloid tip.

TABLE I: Canonical gap distance $\mathcal{D}$ for various 1D anode and cathode geometries

| Panel in Fig. 1 | Anode | Cathode | $\mathcal{D}$ |
|---|---|---|---|
| a | Planar ($x_A$) | Planar ($x_C$) | $x_A - x_C$ |
| b | Concentric Cylinder ($R_A$) | Concentric Cylinder ($R_C$) | $R_C \ln(R_A/R_C)$ |
| c | Eccentric Cylinder ($R_A$) | Eccentric Cylinder ($R_C$) | $R_C \cosh^{-1}\left(\dfrac{R_A^2 + R_C^2 - c^2}{2R_A R_C}\right)$ |
| d | Concentric sphere ($R_A$) | Concentric sphere ($R_C$) | $\dfrac{R_C(R_A - R_C)}{R_A}$ |
| e | Hyperboloid tip ($v_A$) | Hyperboloid tip ($v_C$) | $a \sin^2(v_C) \dfrac{\ln[\tan(v_A/2)]}{\ln[\tan(v_C/2)]}$ |
| f | Planar ($v_A = \pi/2$) | Hyperboloid tip ($v_C$) | $a \sin^2(v_C)[\ln \tan(v_C/2)]$ |
| g | Eccentric sphere ($R_A$) | Eccentric sphere ($R_C$) | $\dfrac{R_C(R_A^3 - R_C^3)(R_A^2 - R_A R_C + cR_C \cos v)}{R_A^2(R_A^3 - R_C^3 + 3cR_C^2 \cos v)}$ |
| h | Infinite plane | Cylinder ($R_C$) | $R_C \ln\left(\dfrac{h + \sqrt{h^2 - R_C^2}}{R_C}\right)$ |
| i | Square ($2h$) | Cylinder ($R_C$) | $R_C \ln\left(\dfrac{1.08h}{R_C}\right)$ |
| j | External cylinder ($R_A$) | External cylinder ($R_C$) | $R_C \left[\ln\left(\dfrac{R_A}{R_C}\right) + \ln\left(\dfrac{c_1 + \sqrt{c_1^2 - R_C^2}}{c_2 + \sqrt{c_2^2 - R_A^2}}\right)\right]$ |
| k | Two intersecting planes ($2\theta$) | Cylinder ($R_C$) | $\dfrac{2\theta R_C}{\pi} \ln\left(\dfrac{2h}{R_C}\right)$ |
| l | Slotted cylinder ($R_A$) | Slotted cylinder ($R_C$) | $R_C \ln\left(\dfrac{R_A}{R_C}\right)$ |
| m | Two planes ($2h$) | Central cylinder ($R_C$) | $R_C \ln\left(\dfrac{4h}{\pi R_C}\right)$ |
| n | Hyperboloid tip ($v_A$) | Tilted hyperboloid tip ($v_C$) | $\left(1 - \frac{\xi^4}{4}\right)^{-1} \left[ a\sin^2(v_C) \left( \ln\left(\frac{\tan\frac{v_A}{2}}{\tan\frac{v_C}{2}}\right) \right) - \frac{1}{2}\ln\left(1 - \frac{\xi^4}{4}\right) \right]$ |
| o | Hyperboloid tip ($v_A$) | Offset hyperboloid tip ($v_C$) | $2a\sin^2(v_C)\ln\left[\frac{\tan(v_A/2)}{\tan(v_C/2)}\right] - 1$ |

## IV. CONCLUSION

We have demonstrated the application of bijective transformations to extend the MGL to 1D nonplanar diodes, extending prior studies that applied these techniques to vacuum SCLCD and the nexus between vacuum SCLCD, the Fowler-Nordheim equation for field emission, and the Richardson-Laue-Dushmann equation for thermal emission [39]. In the process, we have derived a general equation for the canonical gap distance $\mathcal{D}$ in terms of geometric scale factors that replaces the gap distance $D$ in the classical MGL. We further tabulated $\mathcal{D}$ for various complicated geometries that exhibit curvilinear flow. This formulation provides a first step toward more accurately accounting for geometry for the SCLCD in semiconductors and demonstrates the applicability of this approach for any applied current density, including a recent derivation that provided an analytic solution linking the CLL with the MGL [59]. Moreover, the approach used here for deriving $\mathcal{D}$ in Table I holds geometrically for the uniform space-charge limited current density, although the geometric [31] and velocity [35,36] corrections from the CLL differ from those reported here for the SCLCD.

## ACKNOWLEDGMENTS

This work was supported in part by the Joint Directed Energy Transition Office under contract number HQ0642384797, and Sandia National Laboratories (SNL). SNL is a multi-mission laboratory managed and operated by National Technology and Engineering Solutions of Sandia under the U.S. Department of Energy's National Nuclear Security Administration under contract DE-NA0003525. This paper describes objective technical results and analysis. Any subjective views or opinions that might be expressed in the paper do not necessarily represent the views of the U.S. Department of Energy (DOE) or the United States Government. The publisher acknowledges that the U.S. Government retains a non-exclusive, paid-up, irrevocable, world-wide license to publish or reproduce the published form of this written work or allow others to do so, for U.S. Government purposes. The DOE will provide public access to results of federally sponsored research in accordance with the DOE Public Access Plan.

## AUTHOR DECLARATIONS

**Conflict of Interest**

The authors have no conflicts to disclose.

**Author Contributions**

**Allen L. Garner:** Conceptualization (lead); Formal Analysis (supporting); Funding acquisition (lead); Investigation (equal); Methodology (equal); Project Administration (lead); Supervision (lead); Writing – original draft (lead); Writing – review & editing (supporting). **N. R. Sree Harsha:** Conceptualization (supporting), Formal Analysis (equal), Investigation (equal), Methodology (equal), Writing – original draft (supporting), Writing – review and editing (lead). **Amanda M. Loveless:** Formal analysis (equal), Investigation (equal), Methodology (equal), Writing – original draft (supporting) Writing – review & editing (supporting).

## DATA AVAILABILITY

The data that support the findings of this study are available from the corresponding author upon reasonable request.